\documentclass[lettersize,journal]{IEEEtran}

\usepackage[linesnumbered,ruled,vlined]{algorithm2e}
\usepackage{nicefrac}
\usepackage{setspace}
\usepackage{amsthm}

\usepackage{amsmath}%
\usepackage{MnSymbol}%
\usepackage{wasysym}
\usepackage{subcaption}
\usepackage{authblk}
\DeclareMathAlphabet{\pazocal}{OMS}{zplm}{m}{n}

\usepackage[ruled,vlined]{algorithm2e}

\usepackage[mathscr]{euscript}
\usepackage{cite}
\usepackage{psfrag}
\ifCLASSINFOpdf
\usepackage[pdftex]{graphicx}
\graphicspath{{../pdf/}{../jpeg/}}
\DeclareGraphicsExtensions{.pdf,.jpeg,.png}

\else
\usepackage[dvips]{graphicx}
\graphicspath{{../eps/}}
\DeclareGraphicsExtensions{.eps}
\fi
\usepackage{amsmath}
\usepackage{array}

\usepackage{tikz}
\usepackage[utf8]{inputenc}
\usepackage{pgfplots} 
\usepackage{pgfgantt}
\usepackage{pdflscape}
\pgfplotsset{compat=newest} 
\pgfplotsset{plot coordinates/math parser=false}
\usepackage{grffile}
\usetikzlibrary{plotmarks}
\usepackage{amsthm}
\usepackage{amsmath}%
\usepackage{MnSymbol}%
\usepackage{wasysym}
\usepackage{subcaption}
\usepackage[mathscr]{euscript}
\usepackage{cite}
\usepackage{amsfonts}
\usepackage{pgf}
\usepackage{mathtools}
\usepackage{graphicx}
\usepackage{array}
\usepackage{tikz}
\usepackage[utf8]{inputenc}
\usepackage{pgfplots} 
\usepackage{pgfgantt}
\usepackage{pdflscape}
\pgfplotsset{compat=newest} 
\pgfplotsset{plot coordinates/math parser=false}
\usepackage{grffile}
\usetikzlibrary{plotmarks}
\usepackage{tikzscale}
\usepackage{caption}
\usepackage{subcaption}
\usepackage{amsmath}
\usepackage{array}
\usepackage{tikz}
\usepackage[utf8]{inputenc}
\usepackage{pgfplots} 
\usepackage{pgfgantt}
\usepackage{pdflscape}
\pgfplotsset{compat=newest} 
\pgfplotsset{plot coordinates/math parser=false}
\usepackage{grffile}
\usetikzlibrary{plotmarks}
\usepackage[T1]{fontenc}
\usepackage{graphicx}
\usepackage{caption}
\usepackage{overpic}
\usepackage{array}
\usepackage{color}
\usepackage{url}
\usepackage{tikz}
\usepackage[utf8]{inputenc}
\usepackage{pgfplots} 
\usepackage{cite}

\def\CN{\mathcal{C}\mathcal{N}} 

\usepackage[colorlinks=true,bookmarks=false,citecolor=blue,urlcolor=blue]{hyperref}  

\usepackage{xcolor,cite,etoolbox}
\IEEEoverridecommandlockouts

\begin{document}

	\title{Joint Sum Rate and Blocklength Optimization in RIS-aided Short Packet URLLC Systems}

   	\author{Ramin~Hashemi,
	Samad Ali,
	Nurul Huda Mahmood, 
	  and Matti Latva-aho
	  
		
	\vspace{-9mm}

	\thanks{The authors are with the Centre for Wireless Communications (CWC), University of Oulu, 90014 Oulu, Finland. e-mails: (\{ramin.hashemi,  samad.ali, nurulhuda.mahmood,  matti.latva-aho\}@oulu.fi). This research has been supported by the Academy of Finland, 6G Flagship program under Grant 346208.
	
		}
		}
	
	\maketitle

	\begin{abstract}
		In this paper, a multi-objective optimization problem (MOOP) is proposed for maximizing the achievable finite blocklength (FBL) rate while minimizing the utilized channel blocklengths (CBLs) in a reconfigurable intelligent surface (RIS)-assisted short packet communication system. The formulated MOOP has two objective functions namely maximizing the total FBL rate with a target error probability, and minimizing the total utilized CBLs which is directly proportional to the transmission duration. The considered MOOP variables are the base station (BS) transmit power, number of CBLs, and passive beamforming at the RIS. Since the proposed non-convex problem is intractable to solve, the Tchebyshev method is invoked to transform it into a single-objective OP, then the alternating optimization (AO) technique is employed to iteratively obtain optimized parameters in three main sub-problems. The numerical results show a fundamental trade-off between maximizing the achievable rate in the FBL regime and reducing the transmission duration. Also, the applicability of RIS technology is emphasized in reducing the utilized CBLs while increasing the achievable rate significantly. 
	\end{abstract}
	
	
	\IEEEpeerreviewmaketitle
	
    \vspace{-4mm}
	\section{Introduction}
	\bstctlcite{IEEEexample:BSTcontrol}
    Ultra-reliable and low-latency communication (URLLC) has been envisioned to require more stringent key performance indicators (KPIs) in the future 6th generations of wireless communications (6G) \cite{Aceto2019}. Towards this end, reliability in the order of $1-10^{-9}$, latency around $0.1-1$ ms round-trip time, e.g., for industrial control networks, and also high amount of exchanged information bits due to the increased number of sensors/actuators are envisioned \cite{MTCwhitePaper2020,RanjhaAgriculture}. Usually, URLLC packets have short lengths, and the Shannon capacity formula cannot represent the mutual information between transmitted and received bits as the coding is assumed to be performed over an infinite channel blocklength (CBL). Thus, a penalty term exists when only short packets are transmitted in the finite blocklength (FBL) regime \cite{Polyanskiy2010}. However, random propagation nature of wireless channels can have detrimental effects on the performance of URLLC systems resulting in either unexpected delays due to re-transmissions or data rate reductions.

     Reconfigurable intelligent surface (RIS) is a promising technology to control and re-engineer the wireless channel so that the received signals have the desired property \cite{DiRenzo2020}. A RIS is a meta-surface consisting of large and low-cost passive reflecting elements integrated on a planar surface that is reconfigured via a controller such as base stations (BSs). Since there is no processing of signals at the RIS, only the phase shift elements must be assigned optimally to enhance the received signal power \cite{DiRenzo2020}. Also, the control overhead for the RIS is sufficiently lower due to fast backhaul link from BS, and configuring the elements' phase with relatively fast PIN diodes within micro-seconds meets the URLLC demands. And, unlike conventional relays, there is no delay at the RIS as no conversion is performed from the analog to digital domain via radio frequency chains. Moreover, employing RIS supplies the stringent latency and reliability requirements to enhance the quality of URLLC transmission especially when the direct links are blocked. Hence, RIS can have a key role in wireless systems considering URLLC requirements \cite{Ren2022,Hashemi2021a}.

     To date, a number of works have analyzed resource allocation in URLLC systems under FBL regime, e.g., in \cite{Haghifam2017,Ranjha2020,Ren2020_CBL_opt,tehrani2021resource,Ranjha2022Laser}. A seminal work in \cite{Haghifam2017} studied a multi-objective approach to maximize the total FBL rate while minimizing the largest target error probability by designing transmit power of BS, and the target error probabilities for users in a noise-limited regime. An optimization problem (OP) to allocate the utilized CBLs, RIS passive beamforming, and unmanned aerial vehicle (UAV) path planning was proposed in \cite{Ranjha2020}. The scenario was single-user case and the formulated OP was a single-objective problem to minimize the decoding error probability. In \cite{Ren2020_CBL_opt} some tight bounds for transmitting power and allocated CBLs were derived in a URLLC system by considering energy and latency constraints. The authors showed that the relay-assisted networks outperformed direct transmission schemes significantly, though the RIS technology was not the main study topic. Also, \cite{tehrani2021resource} studied minimizing the BS transmit power in a RIS-assisted multiple-input-single-output (MISO)-URLLC secure system to design artificial noise, BS beamformer and RIS passive beamforming where the QoS secrecy rate constraint was also taken into account. A CBL allocation, UAV path planning, and transmit power control resource allocation OP was proposed in \cite{Ranjha2022Laser} to facilitate URLLC in UAV-assisted systems utilizing short packets.

    In this work, we shed some light on optimizing the transmit power, CBL allocation, and RIS passive beamforming by formulating a multi-objective OP where the objectives are to maximize the total FBL rate among users and minimize the leveraged CBLs. Since the transmission duration is directly proportional to the utilized CBLs, our proposed multi-objective approach explores the trade-off between maximizing the achievable FBL rate while reducing the transmission duration. In addition, the potential applicability of the RIS technology to provide an ultra-reliable link while preserving or reducing the transmission duration is discussed.

	In this paper, $\textbf{h} \sim \CN(\textbf{0}_{N\times 1},\textbf{C}_{N\times N})$ denotes circularly-symmetric (central) complex normal distribution vector with zero mean $\textbf{0}_{N\times 1}$ and covariance matrix $\textbf{C}$. The operations $\mathbb{E}[\cdot]$, $[\cdot]^\text{H}$, and $[\cdot]^\text{T}$ denote the statistical expectation, conjugate transpose of a matrix or vector, and the transpose, respectively. $\nabla_\textbf{x}f(\textbf{x})$ indicates the gradient vector of function $f(\textbf{x})$.

	\vspace{-3mm}
	\section{System Model}
	\label{sys_modl}
\begin{figure}[t]
		\centering
		\includegraphics[trim = 9cm 6.5cm 9cm 6cm,scale=0.4]{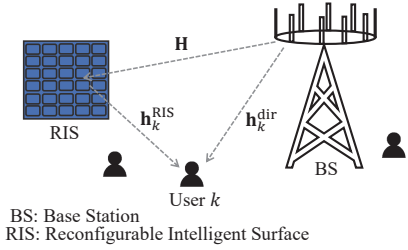}
		\caption{The considered system model.}
		\label{fig:sys_model}
    \end{figure}

	Consider the downlink (DL) of an RIS-assisted wireless network in a multi-user scenario which consists of a BS with $B$ antennas surrounded by $K$ users as shown in Fig. \ref{fig:sys_model}. The RIS has $N$ phase shift elements. The channel between BS and the RIS is defined as $ \textbf{H} = \sqrt{\frac{\zeta^\text{BS}}{\zeta^\text{BS}+1}}\overline{\textbf{H}}+ \sqrt{\frac{{\beta^{\text{BS}}}}{\zeta^\text{BS}+1}}\tilde{\textbf{H}}$ where $\zeta^\text{BS}$ is the Rician factor determining the proportion of line-of-sight (LoS) to non-LoS (NLoS) channel, $\beta^{\text{BS}}$ denotes the path loss coefficient, and each element in $\tilde{\textbf{H}}$ is distributed as an independent and identical distribution (i.i.d.) Gaussian random variable with $\sim\CN(0,1)$. Considering uniform square planar array model, the LoS channel $\overline{\textbf{H}} \in \mathbb{C}^{N\times B}$ is modeled as $\overline{\textbf{H}}=\sqrt{\beta^{\text{BS}}}\textbf{a}_N(\phi_t^a,\phi_t^e)\textbf{a}_B^{\text{H}}(\psi_r^a,\psi_r^e)$ where $\phi_t^a,\phi_t^e$ are the azimuth and elevation angles of arrival (AoA), respectively. Also, $\psi_r^a,\psi_r^e$ are the azimuth and elevation angles of departure (AoD), respectively. Also, $\text{a}_Q(\phi_1,\phi_2)$ is defined as \cite{Zhi2021}
    \begin{flalign}
        \text{a}_Q(\phi_1,\phi_2) \coloneqq & \Big[1,...,e^{\mathrm{j}2\pi \frac{d}{\lambda}\left(n\sin\phi_2\sin\phi_1+n\cos\phi_2\right)},... \\ \nonumber  & ,e^{\mathrm{j}2\pi \frac{d}{\lambda}\left((\sqrt{Q}-1)\sin\phi_2\sin\phi_1+(\sqrt{Q}-1)\cos\phi_2\right)}\Big]^{\text{T}},
    \end{flalign}
    where $\lambda$ is the operating wavelength, and $d$ is the antenna/element spacing. 
    \setlength{\textfloatsep}{1pt}
    
    Similarly, the channel response between the RIS and the user $k$ is shown as $\textbf{h}^{\text{RIS}}_k =\sqrt{\frac{\zeta^\text{RIS}_k}{\zeta^\text{RIS}_k+1}}\overline{\textbf{h}}^{\text{RIS}}_k+\sqrt{\frac{\beta^{\text{RIS}}_k}{\zeta^\text{RIS}_k+1}}\tilde{\textbf{h}}^{\text{RIS}}_k$ where $\tilde{\textbf{h}}^{\text{RIS}}_k\sim\CN(\textbf{0}_{N\times1},\textbf{I}_N)$ in which $\textbf{I}_N$ is an identify matrix of size $N$, LoS channel $\overline{\textbf{h}}^{\text{RIS}}_k = \sqrt{\beta^{\text{RIS}}_k}\textbf{a}_{N}(\varphi_k^a,\varphi_{k}^e)$ for $\forall k \in \mathcal{K}$, and Rician factor $\zeta^\text{RIS}_k$ with path loss $\beta^{\text{RIS}}_k$. Also, $\varphi_k^a,\varphi_{k}^e$ are the azimuth and elevation AoDs, respectively. The direct channel between BS and a typical user $k$ is modeled by $\textbf{h}^{\text{dir}}_k=\sqrt{\frac{\zeta_k^{\text{dir}}}{\zeta_k^{\text{dir}}+1}}\overline{\textbf{h}}^{\text{dir}}_k+ \sqrt{\frac{\beta^{\text{dir}}_k}{\zeta_k^{\text{dir}}+1}}\tilde{\textbf{h}}^{\text{dir}}_k$ where $\tilde{\textbf{h}}^{\text{dir}}_k\sim\CN(\textbf{0}_{B\times1},\textbf{I}_B)$ with factor $\zeta_k^{\text{dir}}$ and LoS component $\overline{\textbf{h}}^{\text{dir}}_k = \sqrt{\beta^{\text{dir}}_k}\textbf{a}_{B}(\tilde{\varphi}_k^a,\tilde{\varphi}_{k}^e)$ defined similarly as for the RIS-users channels where $\tilde{\varphi}_k^a,\tilde{\varphi}_{k}^e$ are the azimuth and elevation AoDs, respectively.

	The received signal at the user $k$ in time slot $t$ is given by
	\begin{flalign} \label{received_signal}
	    y_k[t] = & \overset{\text{User $k$ signal}}{\overbrace{\left({\textbf{h}_k^{\text{dir}}}^{\text{H}} + {\textbf{h}_k^{\text{RIS}}}^\text{H}\boldsymbol{\Theta}\textbf{H}\right) \textbf{w}_k s_k[t]}} \\ \nonumber & + \underset{\text{Interference and noise signal}}{\underbrace{\textstyle\sum_{k'=1, k'\neq k}^{K} \left({\textbf{h}_k^{\text{dir}}}^{\text{H}}  + {\textbf{h}_k^{\text{RIS}}}^\text{H}\boldsymbol{\Theta}\textbf{H}\right) \textbf{w}_{k'} s_{k'}[t] +  n_k[t]}},
	\end{flalign}
	where $s_k[t]$ is the transmitted symbol from the BS such that $\mathbb{E}[|s_k[t]|^2] = p_k$ in which $p_k$ is the transmit power allocated for user $k$, $\textbf{w}_k$ is the precoding vector applied to user $k$ at the BS, and $n_k[t]$ is the additive white Gaussian noise with $\mathbb{E}[|n_k[t]|^2] = \sigma^2 $ in which $\sigma^2=N_0 W$ where $N_0$, $W$ are the noise spectral density and the system bandwidth, respectively. For the sake of simplicity let us denote the overall channel response between BS and the user $k$ as $\textbf{h}_k={\textbf{h}_k^{\text{dir}}}^{\text{H}}  + {\textbf{h}_k^{\text{RIS}}}^\text{H}\boldsymbol{\Theta}\textbf{H}$. The complex reconfiguration matrix $\boldsymbol{\Theta}$ indicates the phase shift and the amplitude attenuation of RIS which is 
	\begin{flalign}
	    \boldsymbol{\Theta}^{N \times N} = \text{diag}(\boldsymbol{\theta}),
	    \enskip \theta_n =  e^{j\phi_n}, \enskip \phi_n \in  [-\pi,\pi), \enskip \forall n \in \mathcal{N}
	\end{flalign}
	where $\boldsymbol{\theta}=[\theta_1,  \theta_2,..., \theta_N]^{\text{H}}$, and $\text{diag}(\cdot)$ is the diagonalized matrix for a given vector. Based on \eqref{received_signal} the signal-to-interference-plus-noise ratio (SINR) of the user $k$ is given by
    \begin{flalign}
      \gamma_k = \frac{p_k | d_{k,k} + \boldsymbol{\theta}^{\text{H}}\textbf{r}_{k,k}   |^2}{\sum_{k'=1,k'\neq k}^{K} p_{k'} | d_{k',k} + \boldsymbol{\theta}^{\text{H}}\textbf{r}_{k',k}   |^2 + \sigma^2},
    \end{flalign}
    where $d_{k',k} = {\textbf{h}_k^{\text{dir}}}^{\text{H}}\textbf{w}_{k'}$, $\textbf{r}_{k',k}={\tilde{\textbf{H}}_k}\textbf{w}_{k'}$, and  ${\tilde{\textbf{H}}_k}=\text{diag}({\textbf{h}_k^{\text{RIS}}}^\text{H})\textbf{H}$. For simplicity let us define $\mathcal{I}_{k}(\textbf{p},\boldsymbol{\theta})=\sum_{k'=1,k'\neq k}^{K} p_{k'} | d_{k',k} + \boldsymbol{\theta}^{\text{H}}\textbf{r}_{k',k}   |^2 + \sigma^2$ in which $\textbf{p}=[p_1,...,p_K]$. The BS applies maximal ratio transmission (MRT) precoding to send the desired signal to users, i.e., $\textbf{w}_k^{\text{MRT}}=\frac{\textbf{h}_k}{\lVert\textbf{h}_k\rVert}$ $\forall k$\footnote{In this work, perfect channel state information (CSI) is assumed to be available to assess the system's upperbound performance. Also, Monte Carlo simulations for imperfect CSI are provided in Section \ref{numerical_results} for comparison.}.

	In FBL regime the number of information bits that can be transmitted through $m_k$ channel uses is given by \cite{Polyanskiy2010}
	\begin{flalign}
        L_k =  m_k\text{C}(\gamma_k) - Q^{-1}(\varepsilon_k)\sqrt{m_k\text{V}(\gamma_k)} + \log_2(m_k),
         \label{achievable_rate_urllc}
    \end{flalign}
    where $\mathcal{O}\left(\log_2(m_k)\right)\approx\log_2(m_k)$ and $\text{C}(\gamma_k) = \log_2(1+\gamma_k)$ is the Shannon capacity formula and $\varepsilon_k$ is defined as the target error probability for user $k$ and $Q^{-1}(\cdot)$ is the inverse of the Q-function\footnote{The Q-function is defined as $Q(x) = \frac{1}{\sqrt{2\pi}}\int_{x}^{\infty}e^{-\nu^2/2}d\nu$.}. The channel dispersion is defined $\text{V}(\gamma_k) = \frac{1}{(\ln2)^2} \big( 1- \frac{1}{(1+\gamma_k)^2} \big)$. We assume $\text{V}(\gamma)\approx \frac{1}{(\ln2)^2}$\footnote{Typically, the value of $\text{V}(\gamma)\geq \frac{0.99}{(\ln2)^2}$ for $\gamma \geq 20$ dB \cite{Ranjha2022Laser,Ren2020_CBL_opt}.}. Solving \eqref{achievable_rate_urllc} to find the decoding error probability $\varepsilon_k$ at the user $k$ yields $\varepsilon_k = Q\left(f(\gamma_k,m_k,L_k)\right)$ where $ f(\gamma_k,m_k,L_k) = \sqrt{\frac{m_k}{V(\gamma_k)}}(\log_2(1+\gamma_k)-\frac{L_k}{m_k})$. Also, when the CBL $m_k$ asymptotically goes infinity the FBL rate simplifies to $\lim_{m_k \rightarrow \infty} \frac{L_k}{m_k} = \log_2\left(1+\gamma_k\right)$ which is the conventional Shannon capacity formula. Let us define the CBL vector as $\textbf{m}=[m_1,...,m_K]$.
	
	\vspace{-3mm}
    \section{Problem Formulation and solution approach}
    \label{problem_form}
    Maximizing total transmitted information bits to increase the data rate while guaranteeing a target error probability, and minimizing the number of used CBLs are essential in URLLC systems. Thus, we formulate the following multi-objective OP:
	\begin{subequations}
	\begin{flalign} \label{P1obj}
		\textbf{P1 }&
		\begin{cases}
			\displaystyle
			\max_{\textbf{p},\textbf{m},\boldsymbol{\theta}} \enskip L_{\text{total}}(\textbf{p},\textbf{m},\boldsymbol{\theta}) = \textstyle\sum_{k=1}^{K}L_k \\
			\displaystyle
			\min_{\textbf{p},\textbf{m},\boldsymbol{\theta}} \enskip
			m_{\text{total}}=\textstyle\sum_{k=1}^{K}m_k
		\end{cases}
		\\ 
		\text{\textbf{s.t.}} \quad 
		& \text{C$_\text{1}$: } \textstyle\sum_{k=1}^{K}m_k \leq M, \enskip m_k \geq m_k^{\text{min}}, \enskip  \forall k \in \mathscr{K}, \\
		& \text{C$_\text{2}$: } \textstyle\sum_{k=1}^{K} p_k \leq p^{\text{total}}, \enskip  \text{C$_\text{3}$: }
		|\theta_n|=1, \enskip n \in \mathcal{N}.  
	\end{flalign} 
	\end{subequations}
  where the objectives are to jointly maximize the total number of information bits that are transmitted to the users while minimizing the total utilized CBLs to reduce the transmission duration. The optimization variables are CBL variables $\textbf{m}$, BS transmit power vector $\textbf{p}$, and RIS phase shift setting. The constraint C$_\text{1}$ implies that each user's allocated CBL should be at least greater than $m_k^{\text{min}}$ and the total available CBLs should be less than a maximum value. The constraints C$_\text{2}$, and C$_\text{3}$ are the BS maximum transmit power, and the RIS phase shift constraint, respectively. Note that \textbf{P1} belongs to a class of nonlinear and non-convex OP which is intractable to solve. Next, we study the transformation of the \textbf{P1} into an equivalent single-objective OP and proposing the solution approach.
    
    \vspace{-3mm}
    \subsection{Problem Transformation}
    First, we apply Tchebyshev method \cite{Marler2004} to transform the problem into a single-objective equivalent OP as follows:
	\begin{flalign} \label{P2obj}
		\textbf{P2 }&
		\min_{\textbf{p},\textbf{m},\boldsymbol{\theta},\mu} \enskip \mu
		\\
		\text{\textbf{s.t.}} \quad \nonumber
		&\text{C$_\text{1}$--C$_\text{3}$}, \tilde{\text{C}}\text{$_\text{1}$: }{\alpha}/{L_{\text{total}}^*}(L_{\text{total}}^* - L_{\text{total}}(\textbf{p},\textbf{m},\boldsymbol{\theta})) \leq \mu, \\\nonumber
		& \tilde{\text{C}}\text{$_\text{2}$: }({1-\alpha})/{m_{\text{total}}^*}(m_{\text{total}}-m_{\text{total}}^*) \leq \mu. \nonumber 
	\end{flalign}
	where $\mu$ is an auxiliary parameter, $0\leq \alpha \leq 1$ is the non-negative weight which is set by a decision maker. Additionally, the utopia points are found as \cite{Marler2004}
	\begin{flalign} 
		L_{\text{total}}^*= & 
		\max_{\textbf{p},\textbf{m},\boldsymbol{\theta}} \enskip L_{\text{total}} \enskip \text{and }     m_{\text{total}}^* = \min_{\textbf{p},\textbf{m},\boldsymbol{\theta}} \enskip m_{\text{total}} \enskip 
		\text{\textbf{s.t.}} \enskip \text{C$_\text{1}$--C$_\text{3}$}. 
	\end{flalign} 
	Clearly $m_{\text{total}}^*=\sum_{k=1}^{K}m_k^{\text{min}}$ since there is no constraint except C$_{1}$ when the problem is to be solved only for $\textbf{m}$. Also to compute $L_{\text{total}}^*$ a similar method as explained in the following section can be employed as a special case without CBL objective. Herein, we employ alternating optimization technique to divide the problem \textbf{P2} into three sub-problems which will be discussed separately in the subsequent sections.

    \vspace{-3mm}
	\subsection{Transmit Power Allocation}
	By fixing  $\textbf{m}=\textbf{m}^i$ and $\boldsymbol{\theta}=\boldsymbol{\theta}^i$ which are feasible starting points, the sub-problem of finding $\textbf{p}$ is written as
	\begin{flalign}
        \min_{\textbf{p},\mu} \enskip \mu \enskip\text{\textbf{s.t.}} \enskip \nonumber
		&\text{C$_\text{2}$}, \tilde{\text{C}}\text{$_\text{1}$: }{\alpha}/{L_{\text{total}}^*}(L_{\text{total}}^* - L_{\text{total}}(\textbf{p},\textbf{m}^i,\boldsymbol{\theta}^i)) \leq \mu, 
	\end{flalign}
	which is mathematically equivalent to:
	\begin{flalign} \label{P22obj_equiv}
		\max_{\textbf{p}} \enskip &  L_{\text{total}}(\textbf{p},\textbf{m}^i,\boldsymbol{\theta}^i), \quad \text{\textbf{s.t.}} \quad 
		 \text{C$_\text{2}$: } \textstyle\sum_{k=1}^{K} p_k \leq p^{\text{total}},  
	\end{flalign}
	In general, $L_{\text{total}}(\textbf{p},\textbf{m},\boldsymbol{\theta})$ can be rewritten as
	\begin{flalign}
	    &L_{\text{total}}(\textbf{p},\textbf{m},\boldsymbol{\theta})  =  \tilde{L}_t^{+}(\textbf{p},\textbf{m},\boldsymbol{\theta})-\tilde{L}_t^{-}(\textbf{p},\textbf{m},\boldsymbol{\theta}), \\ \nonumber 
        &\tilde{L}_t^{+}(\textbf{p},\textbf{m},\boldsymbol{\theta}) = \textstyle\sum_{k=1}^{K}\big[m_k\log_2\left(\mathcal{I}_{k}(\textbf{p},\boldsymbol{\theta}) + p_k|d_{k,k} + {\boldsymbol{\theta}}^{\text{H}}\textbf{r}_{k,k}   |^2\right) \\ \nonumber & \quad \quad \quad  \quad \quad\quad\quad\quad +\log_2(m_k)\big], \\ \nonumber
        &\tilde{L}_t^{-}(\textbf{p},\textbf{m},\boldsymbol{\theta})=\textstyle\sum_{k=1}^{K}\left[m_k\log_2\left(\mathcal{I}_{k}(\textbf{p},\boldsymbol{\theta})\right)+ \frac{\sqrt{m_k}}{\ln2}Q^{-1}(\varepsilon_k)\right].\nonumber 	    
	\end{flalign}

	It can be proved that $ \tilde{L}_t^{+}(\textbf{p},\textbf{m}^i,\boldsymbol{\theta}^i)$ and $ \tilde{L}_t^{-}(\textbf{p},\textbf{m}^i,\boldsymbol{\theta}^i)$ are concave functions in terms of $\textbf{p}$ \cite{boyd2004convex}, and $L_{\text{total}}(\textbf{p},\textbf{m}^i,\boldsymbol{\theta}^i)$ is difference of concave (DC) functions. Hence, we can apply successive convex approximation (SCA) method by writing first order Taylor series expansion around $\textbf{p}^{i}$ which yields $ L_{\text{total}}(\textbf{p},\textbf{m}^i,\boldsymbol{\theta}^i) \leq  \tilde{L}_t(\textbf{p},\textbf{m}^i,\boldsymbol{\theta}^i)$ where 
	\begin{flalign}
	   \tilde{L}_t(\textbf{p},\textbf{m}^i,\boldsymbol{\theta}^i)  & =   \tilde{L}_t^{+}(\textbf{p},\textbf{m}^i,\boldsymbol{\theta}^i) - \tilde{L}_t^{-}(\textbf{p}^i,\textbf{m}^i,\boldsymbol{\theta}^i) \nonumber \\  & \quad \quad-  \nabla_{\textbf{p}} \tilde{L}_t^{-}(\textbf{p},\textbf{m}^i,\boldsymbol{\theta}^i)^{\text{T}}\left(\textbf{p}-\textbf{p}^i\right),
	\end{flalign}
	thus, the following convex problem can be solved iteratively
	\begin{flalign} 
		\text{\textbf{P2.1}: } \max_{\textbf{p}} \enskip & \tilde{L}_t(\textbf{p},\textbf{m}^i,\boldsymbol{\theta}^i), \quad	\text{\textbf{s.t.}} \quad \nonumber
		 \text{C$_\text{2}$: }\textstyle \sum_{k=1}^{K} p_k \leq p^{\text{total}}, \nonumber 
	\end{flalign}
	until some convergence criteria to find optimized $\textbf{p}^{\text{opt}}$. It is worth noting that the CVX tool \cite{grant2008cvx} can also be used here. 
	
	\vspace{-3mm}
	\subsection{Channel Blocklength Optimization}
	By fixing $\textbf{p}=\textbf{p}^i$ and $\boldsymbol{\theta}=\boldsymbol{\theta}^i$ in which $\textbf{p}^i$ and $\boldsymbol{\theta}^i$ are feasible points, the optimized values for CBL vector $\textbf{m}$ should be determined. The following sub-problem should be solved:
	\begin{flalign} \label{P21obj}
		\min_{\textbf{m},\mu} \enskip & \mu \enskip 
		\text{\textbf{s.t.}} \enskip 
		\text{C$_\text{1}$}, \tilde{\text{C}}\text{$_\text{2}$}, \tilde{\text{C}}\text{$_\text{1}$: } L_{\text{total}}(\textbf{p}^i,\textbf{m},\boldsymbol{\theta}^i) \leq L_{\text{total}}^*(\frac{\mu }{\alpha}-1), 
	\end{flalign}
	similar to the previous section $L_{\text{total}}(\textbf{p}^i,\textbf{m},\boldsymbol{\theta}^i) = \tilde{L}_t^{+}(\textbf{p}^i,\textbf{m},\boldsymbol{\theta}^i)-\tilde{L}_t^{-}(\textbf{p}^i,\textbf{m},\boldsymbol{\theta}^i)$ is a DC function, thus applying SCA method to approximate $\tilde{L}_t^{-}(\textbf{p}^i,\textbf{m},\boldsymbol{\theta}^i)$ yields
    \begin{flalign}
	    \tilde{L}_t^{-}(\textbf{p}^i,\textbf{m},\boldsymbol{\theta}^i)\leq  \underset{\bar{L}'_t(\textbf{p}^i,\textbf{m},\boldsymbol{\theta}^i)}{\underbrace{\textstyle\sum_{k=1}^{K}Q^{-1}(\varepsilon_k)\frac{\sqrt{m_k^i\text{V}(\gamma_k^i)}}{2}\left(1 +\frac{m_k}{m_k^{i}}\right)}}, \label{upp_bound}
	\end{flalign}
	Finally, the CBL allocation sub-problem is given by
	\begin{flalign} 
		 \text{\textbf{P2.2}: } & \min_{\textbf{m},\mu} \enskip  \mu 
		\\
		\text{\textbf{s.t.}}  \enskip  \text{C$_\text{1}$}, & \tilde{\text{C}}\text{$_\text{2}$}, \tilde{\text{C}}\text{$_\text{1}$: } \bar{L}'_t(\textbf{p}^i,\textbf{m},\boldsymbol{\theta}^i)- \tilde{L}_t^{+}(\textbf{p}^i,\textbf{m},\boldsymbol{\theta}^i) \leq L_{\text{total}}^*(\frac{\mu }{\alpha}-1), \nonumber 
	\end{flalign}	
	since the problem \textbf{P2.2} is convex, it can be solved iteratively until convergence by leveraging CVX \cite{grant2008cvx} after replacing the upperbound in \eqref{upp_bound} starting from the feasible point $\textbf{m}^i$.
	
	\vspace{-5mm}
	\subsection{Passive Beamforming Design}
	The phase shift optimization at the RIS is formulated as:
	\begin{flalign}
		\nonumber \min_{\boldsymbol{\theta},\mu} \enskip & \mu \enskip
		\text{\textbf{s.t.}} \enskip \text{C$_\text{3}$}, \tilde{\text{C}}\text{$_\text{1}$: }{\alpha}/{L_{\text{total}}^*}(L_{\text{total}}^* - L_{\text{total}}(\textbf{p}^i,\textbf{m}^i,\boldsymbol{\theta})) \leq \mu,
	\end{flalign}
	after removing excess terms that do not depend on $\boldsymbol{\theta}$ the above problem is equivalent to:
	\begin{flalign}
		\max_{\boldsymbol{\theta}} \enskip &\textstyle \sum_{k=1}^{K}m_k^i\log_2(1+\gamma_k), \quad \text{\textbf{s.t.}} \quad 
		 |\theta_n|=1, \enskip n \in \mathcal{N}. 
	\end{flalign}
	to solve the above problem we start by writing the equivalent form of the logarithm function inside the objective for $\forall k$ via introducing auxiliary variables $\boldsymbol{\kappa}=[\kappa_1,\kappa_2,...,\kappa_K]$ as
	\begin{flalign}
	    \ln(1+\gamma_k) = \max_{\kappa_k\geq 0} \enskip \ln(1+\kappa_k) - \kappa_k + \frac{\gamma_k(1+\kappa_k)}{1+\gamma_k},
	\end{flalign}
	also, let us define the nominator and denominator of the $\gamma_k=\frac{|A_k(\boldsymbol{\theta})|^2}{B_k(\boldsymbol{\theta})-|A_k(\boldsymbol{\theta})|^2}$ where $A_k(\boldsymbol{\theta})= \sqrt{p_k^i} (d_{k,k} + \boldsymbol{\theta}^{\text{H}}\textbf{r}_{k,k}   )$ and $B_k(\boldsymbol{\theta})=\sum_{k'=1}^{K} p_{k'}^i |d_{k',k} + \boldsymbol{\theta}^{\text{H}}\textbf{r}_{k',k}   |^2 + \sigma^2$. Thus, the original problem is transformed to \cite{Huayan2020}
	\begin{flalign}
   		\max_{\boldsymbol{\theta},\boldsymbol{\kappa}} \underset{g(\boldsymbol{\theta},\boldsymbol{\kappa})}{\underbrace{\sum_{k=1}^{K}m_k^i\left(\ln(1+\kappa_k)-\kappa_k + \frac{|A_k(\boldsymbol{\theta})|^2}{B_k(\boldsymbol{\theta})}(\kappa_k+1)\right)}}, \enskip 
		\text{\textbf{s.t.}} \enskip 
		\text{C$_\text{3}$}. \nonumber  
	\end{flalign}
	The above problem can be tackled in two steps. First, by considering fixed $\boldsymbol{\theta}$ there will be a concave objective function with respect to $\kappa_k$ $\forall k$ and the optimal point can be obtained by calculating $\frac{\partial g(\boldsymbol{\theta},\kappa_k)}{\partial \kappa_k}=0$ which yields $\kappa_k^{\text{opt}}=\gamma_k$ where after substituting in $g(\boldsymbol{\theta},\boldsymbol{\kappa})$, it yields the following problem:
    \begin{flalign}
        \max_{\boldsymbol{\theta}} \enskip \textstyle\sum_{k=1}^{K}m_k^i(1+\kappa_k)\frac{|A_k(\boldsymbol{\theta})|^2}{B_k(\boldsymbol{\theta})},
    \end{flalign}
    which is a fractional programming (FP) and can be cyclically tackled by defining auxiliary variables $\boldsymbol{\xi}=[\xi_1,\xi_2,...,\xi_K]$ that reformulates the objective function as:
    \begin{flalign}
        \max_{\boldsymbol{\theta},\boldsymbol{\xi}}  \sum_{k=1}^{K}\left(2\sqrt{m_k^i(1+\kappa_k)}\mathcal{R}\{\xi^*_kA_k(\boldsymbol{\theta})\}-|\xi_k|^2 B_k(\boldsymbol{\theta})\right),\label{fp_eq}
    \end{flalign}
    For given $\boldsymbol{\theta}$, the optimal $\boldsymbol{\xi}^{\text{opt}}$ is easily obtained by computing the roots of partial derivative of the objective function in \eqref{fp_eq} with respect to $\xi_k$. The final result is given by
    \begin{flalign}
        \xi^{\text{opt}}_k=\frac{\sqrt{p_k^i m_k^i(1+\kappa_k^{\text{opt}})}(d_{k,k} + \boldsymbol{\theta}^{\text{H}}\textbf{r}_{k,k})}{\sum_{k'=1}^{K} p_{k'}^i |d_{k',k} + \boldsymbol{\theta}^{\text{H}}\textbf{r}_{k',k}   |^2 + \sigma^2},
    \end{flalign}
    next, by substituting $\boldsymbol{\xi}^{\text{opt}}=[\xi_1^{\text{opt}},...,\xi_K^{\text{opt}}]$ in \eqref{fp_eq} the OP for $\boldsymbol{\theta}$ is reformulated as
    \begin{flalign}
        \text{\textbf{P2.3}: }\max_{\boldsymbol{\theta}} \enskip & -\boldsymbol{\theta}^{\text{H}}\textbf{Q}\boldsymbol{\theta}+2\mathcal{R}\{\boldsymbol{\theta}^{\text{H}}\textbf{q} \}, \enskip 
		\text{\textbf{s.t.} }  \boldsymbol{\theta}^{\text{H}}\textbf{e}_n\textbf{e}_n^{\text{H}}\boldsymbol{\theta} \leq 1, \enskip n \in \mathcal{N}. \nonumber 
    \end{flalign}
    where $\textbf{e}_n \in \mathbb{R}^{N\times1}$ is a vector where the n-th element is one and zero elsewhere. Moreover, $\textbf{Q}$ and $\textbf{q}$ are given by
    \begin{flalign}
        \textbf{Q} = & \textstyle\sum_{k=1}^{K}|\xi_k^{\text{opt}}|^2\sum_{k'=1}^{K}p_{k'}^i\textbf{r}_{k',k}\textbf{r}_{k',k}^{\text{H}}, \\ \nonumber 
        \textbf{q} = & \sum_{k=1}^{K}\left(\sqrt{p_k^i (1+\kappa_k)m_k^i}(\xi_k^{\text{opt}})^*\textbf{r}_{k,k}-|\xi_k^{\text{opt}}|^2\sum_{k'=1}^{K}p_{k'}^i d_{k',k}^*\textbf{r}_{k',k}\right),
    \end{flalign}
	it can be proved that \textbf{P2.3} is a quadratically constrained quadratic program (QCQP) as the objective and the constraints are quadratic functions in terms of $\boldsymbol{\theta}$. Several QCQP solvers and methods can be leveraged to obtain the optimized values of passive beamforming at the RIS, e.g., CVX\footnote{Several efficient algorithms are available for RIS phase shift design \cite{pan2021overview}, however our aim is to discuss the general idea of rate/CBL trade-off.}. The summarized algorithm to solve \textbf{P2} is provided in Algorithm \ref{Algorithm1}.
	
	\vspace{-3mm}
	\subsection{Complexity Analysis}
	As \textbf{P2.1} is a convex OP with  $K$ variables and one constraint, it can be solved via primal-dual interior-point methods in polynomial time complexity in terms of the number of constraints/variables \cite{boyd2004convex,grant2008cvx}. Thus, to achieve a duality gap precision of $\delta$ in \textbf{P2.1}, the complexity order will be $\mathcal{O}(K^{3.5}\ln(\delta^{-1}) I_1)$ where $I_1$ is the number of iterations. A similar approach can be used for \textbf{P2.2} that has $K+1$ variables and $K+3$ constraints which has $\mathcal{O}\left(K^{3.5}\ln(\delta^{-1})I_2\right)$ where $I_2$ is the number of iterations. In \textbf{P2.3}, $\boldsymbol{\kappa}$/$\boldsymbol{\xi}$, and $\textbf{Q}$ have $\mathcal{O}(2KNM)$, and $\mathcal{O}(K^2N^2)$ computational complexity, respectively. Moreover, solving a QCQP with $N$ complex variables and constraints has $\mathcal{O}(N^{3.5}\ln(\delta^{-1}))$. Thus, the total complexity of \textbf{P2.3} is $\mathcal{O}\left((K^2N^2+2KNM+N^{3.5}\ln(\delta^{-1}))I_3\right)$ where $I_3$ is the number of iterations. Overall, the Algorithm \ref{Algorithm1} has polynomial time complexity.

	\begin{algorithm}[t]
	\small
	\SetAlgoLined 
 	\caption{Joint transmit power, phase shift and CBL allocation.}
 	\label{Algorithm1}
 	\KwIn{Total transmit power, utopia points, $\alpha$, $\varepsilon_k$ $\forall k$.}
 	\KwOut{The optimized values  $\textbf{p}^{\text{opt}}$,  $\textbf{m}^{\text{opt}}$, and $\boldsymbol{\theta}^{\text{opt}}$}
 	\textit{Initialization}: $i=0$, $\textbf{p}^{0}$, $\textbf{m}^{0}$, and $\boldsymbol{\theta}^{0}$\;
 	\While{Convergence}{
 	Solve \textbf{P2.1} given $\textbf{m}^i$ and $\boldsymbol{\theta}^i$ to find optimized $\textbf{p}=\textbf{p}^{i}$\;
   	Solve \textbf{P2.2} given $\textbf{p}^i$ and $\boldsymbol{\theta}^i$ to find optimized $\textbf{m}=\textbf{m}^{i}$\;
  	Solve \textbf{P2.3} given $\textbf{p}^i$ and $\textbf{m}^i$ to find optimized $\boldsymbol{\theta}=\boldsymbol{\theta}^{i}$\;
  	$i = i +1$\;
  	
 	}
 \end{algorithm}
\setlength{\textfloatsep}{0pt}

	\vspace{-2.5mm}
	\section{Numerical Results}
	\label{numerical_results}
	We numerically evaluate the proposed joint CBL, transmit power optimization and passive beamforming in this section. The network topology consists of a BS at centre coordinates and $K=4$ users which are randomly located at $10$ meters distance from the RIS which is positioned at $[200,0]$ in 2-D plane. The large scale path loss for the reflected channel is modeled as $\text{PL(dB)}=-30-22\log_{10}(d[m])$ where $d[m]$ shows the distance between a transmitter and receiver. The path loss model for direct channel is assumed $\text{PL}_{\text{dir}}\text{(dB)}=-33-38\log_{10}(d[m])$. The demonstrated curves are the averaged optimizations' results over $100$ realizations of independent channels' small-scale fading coefficients. To model the imperfect channel state information (CSI), a similar method as in \cite{Huayan2020} is employed where each small-scale fading component $h$ with estimated value of $\hat{h}$ has an error term $e\sim \CN(0,\sigma_e^2)$ such that $h=\hat{h}+e$. The normalized mean square error (MSE) coefficient $\rho={\mathbb{E}[|h-\hat{h}|^2]}/{\mathbb{E}[|\hat{h}|^2]}$ is used to control the CSI uncertainties \cite{Huayan2020}. Table \ref{tab1} shows the considered parameters selected for the network.
	\begin{table}[t]
	\small
		\caption{Simulation parameters.}
		\centering
		\begin{tabular}{ l  l }
			\hline
			Parameter & Default value \\ \hline
			
			Number of users ($K$)  & 4  \\ 
			Number of BS antennas ($B$) & 4 \\
			Number of RIS elements ($N$) & 25 \\
            Noise power density ($N_0$) & -174 dBm/Hz \\ 
            BS maximum transmit power ($p^{\text{total}}$)  & 10 mW  \\
            Target error probability ($\varepsilon_k$ $\forall k$)  & $10^{-6}$  \\
            Tchebyshev coefficient ($\alpha$)  & 0.8 \\
			Maximum available CBL ($M$) & 200 \\ 
			Minimum CBL ($m^{\text{min}}_k$ $\forall k$) & 10 \\ 
		    Rician factors ($\zeta^{\text{BS}}$, $\zeta^{\text{RIS}}_k$, $\zeta^{\text{dir}}_k$  $\forall k$) & 1 \\ 
			Bandwidth ($W$) &  2 MHz \\ 
			Distance from RIS to the BS & 20 m \\
			\hline
		\end{tabular}
		\label{tab1}
	\end{table}
	\setlength{\textfloatsep}{2pt}
	
    \begin{figure*}[t]
     \begin{subfigure}{0.33\linewidth}
 		\centering
 		\includegraphics[trim = 9cm 9cm 9cm 9cm,scale=0.4]{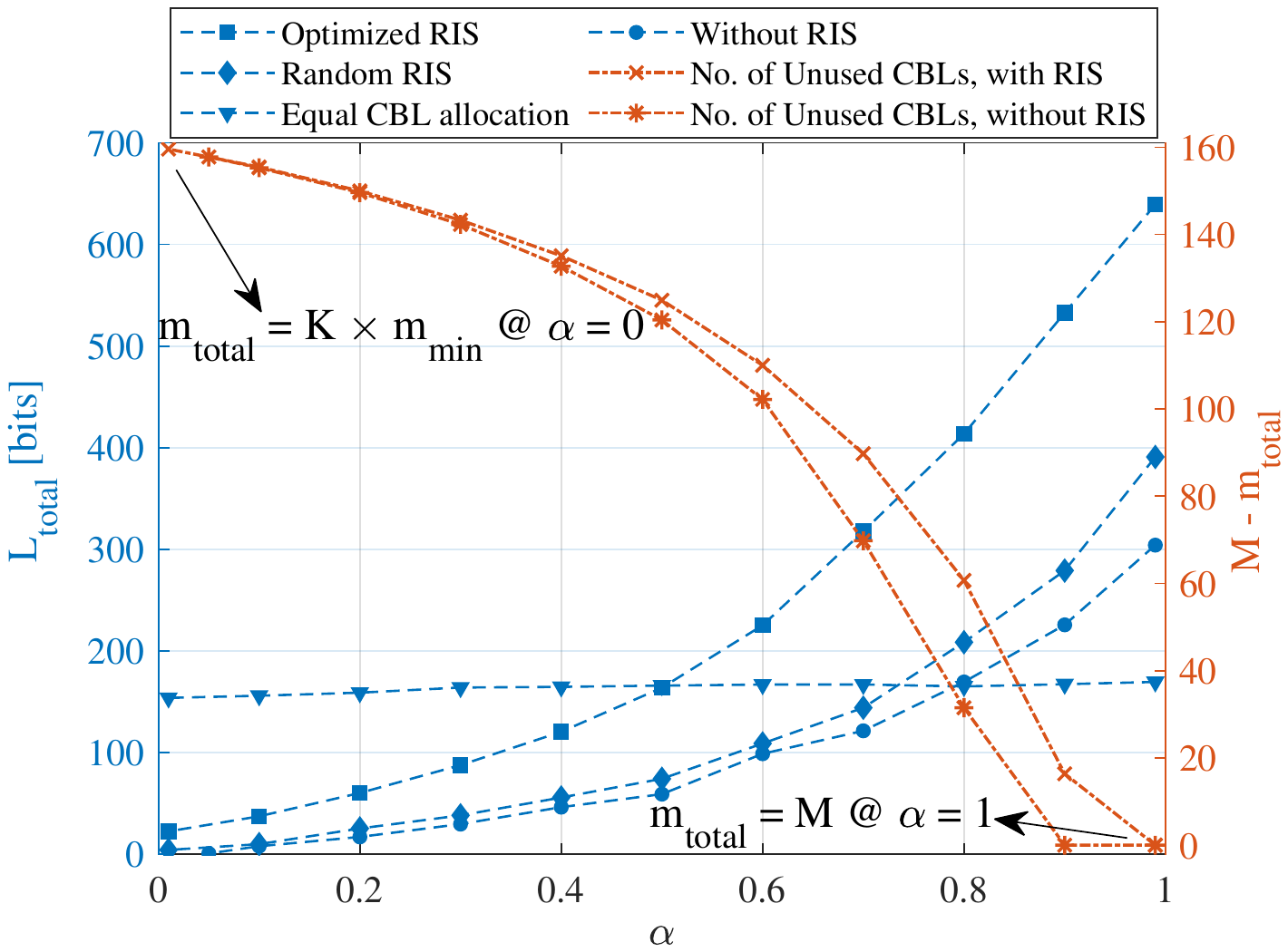}
 		\caption{ }
 		\label{fig:L}
     \end{subfigure}
    \begin{subfigure}{0.33\linewidth}
 		\centering
 		\includegraphics[trim = 9cm 9cm 9cm 9cm,scale=0.4]{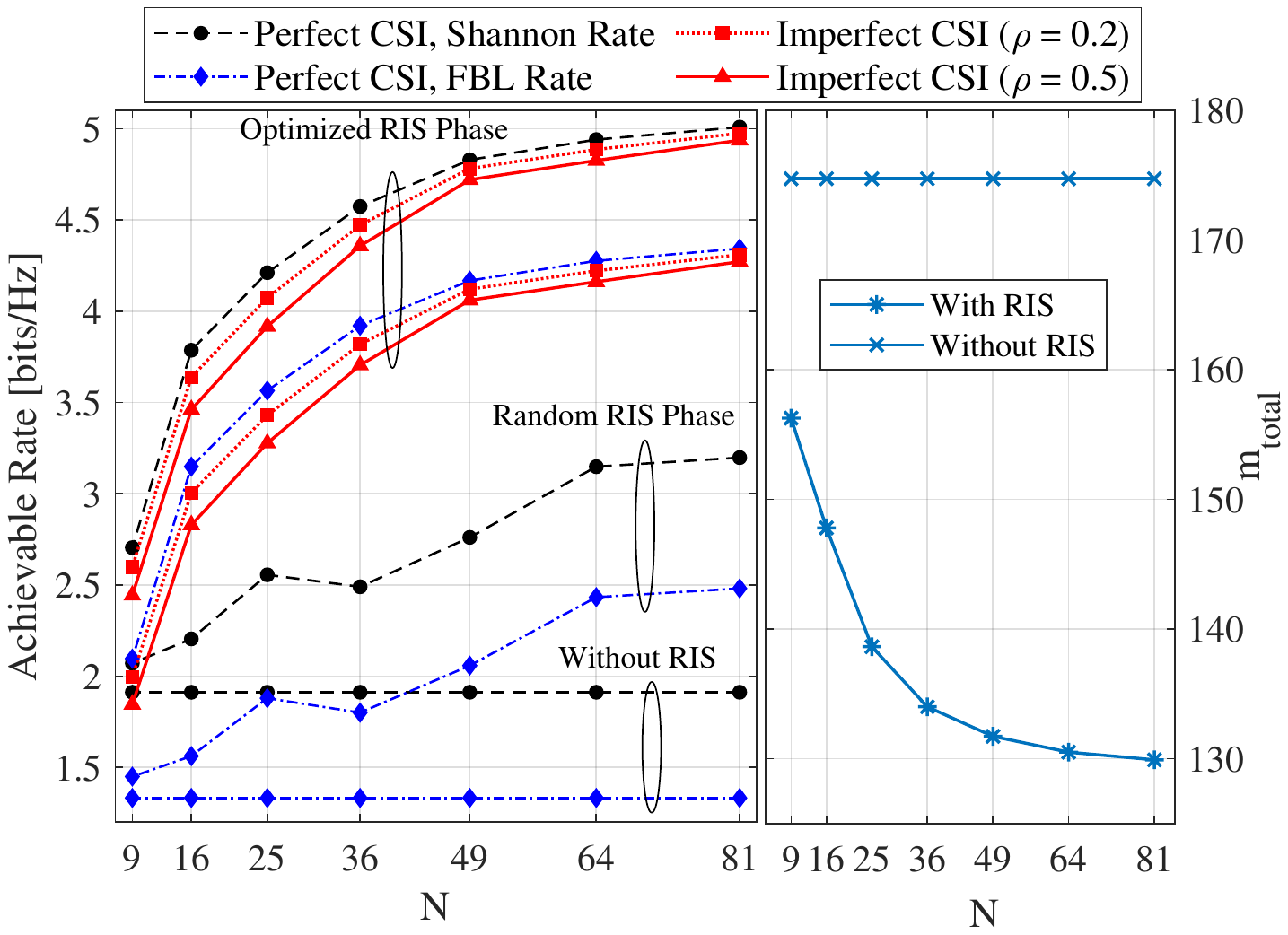}
 		\caption{ }
 		\label{fig:N}
     \end{subfigure}
      \begin{subfigure}{0.33\linewidth}
 		\centering
 		\includegraphics[trim = 9cm 9cm 9cm 9cm,scale=0.4]{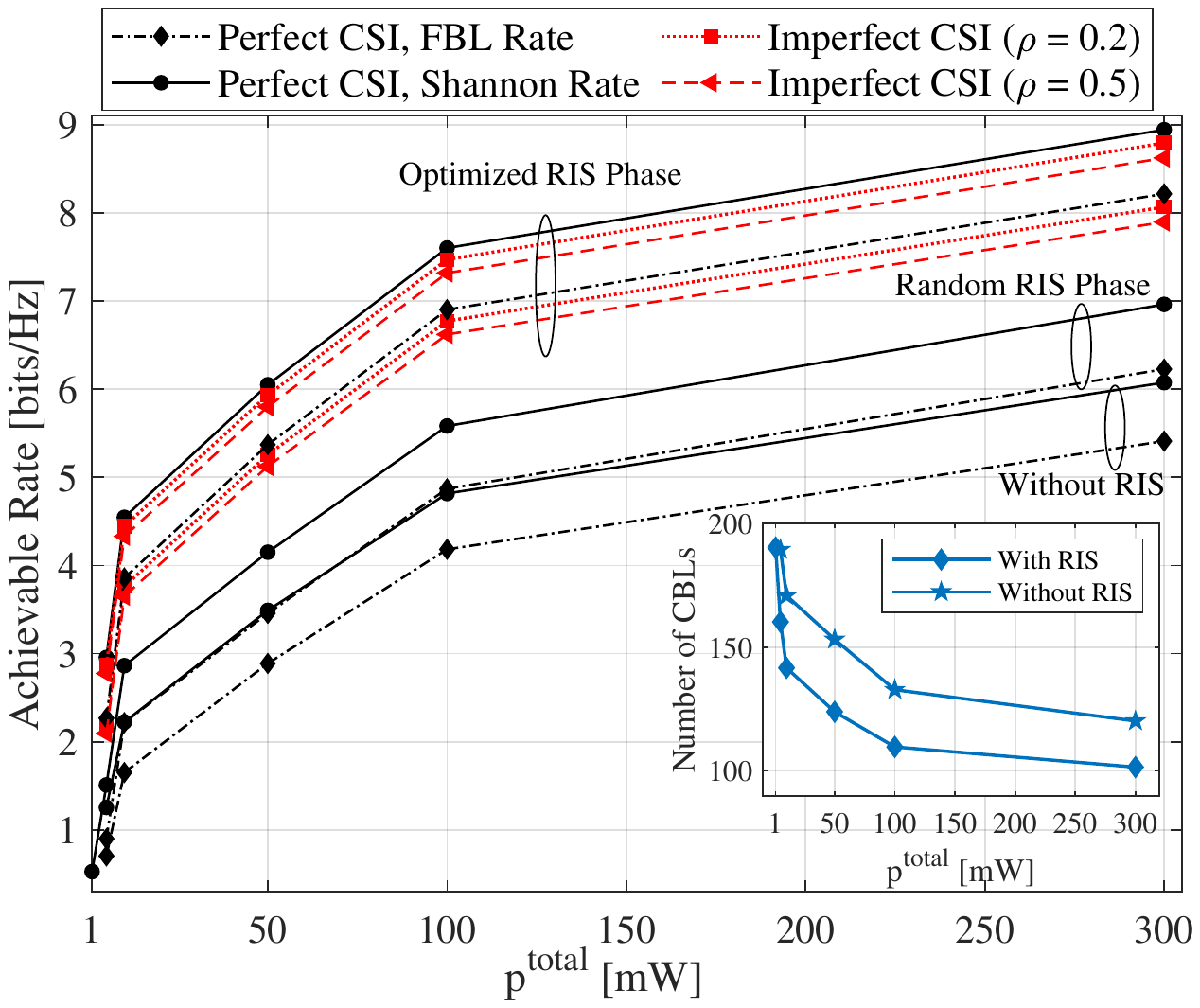}
 		\caption{ }
 		\label{fig:P}
     \end{subfigure}
      \caption{(a): The trade-off illustration between two considered objective functions. (b): The impact of increasing the number of elements at the RIS. (c): The impact of increasing maximum transmit power at the BS on the achievable rate.}
      \label{fig:convergence_fig}
    \end{figure*}
	
    Fig. \ref{fig:L} illustrates the performance of the multi-objective OP in terms of varying $\alpha$ considering several benchmarks as shown. As it is observed from the conflicting objective functions in \textbf{P1} when $\alpha$ is closer to zero, the used CBLs reduce, and in the extreme case, it is equal to $K\times m^{\text{min}}$ meaning that each user is assigned to only $m^{\text{min}}$, and the FBL rate decreases as well. Since $m_{\text{total}}$ increases with respect to $\alpha$, the transmission duration $T$ grows as it is directly impacted by the consumed CBLs with $T=\frac{m}{W}$ where $W$ are the bandwidth, $m$ is the utilized CBLs. Thus, there is a trade-off between the FBL rate and the transmission duration. Note that for higher $\alpha$, the weight value for the OP related to maximizing $L_{\text{total}}$ has higher significance than minimizing the utilized CBL.

    In Fig. \ref{fig:N} the system total FBL rate and utilized CBLs are plotted for a different number of RIS elements. When the number of elements at the RIS increases, the amount of exploited CBLs reduces by 17\%, and the achievable FBL rate rises. Thus, the conflicting property in the objective functions of \textbf{P1} can be compensated by increasing the number of phase shift elements at the RIS. In addition, there is a noticeable gap between having a RIS with optimized phase shifts and performing random passive beamforming at the RIS which is around 48\% meaning that the phase shift optimization results in 48\% enhancement in the total FBL rate when $N=25$.

	Finally, Fig. \ref{fig:P} depicts the performance of the achievable rate, and used CBLs in terms of varying the total transmit power at the BS. As can be seen, for higher maximum transmit power budget, both optimized FBL and Shannon rate increase monotonically. In addition, the utilized number of CBLs reduces by at most around 45\% until $p^{\text{total}}=300$ mW. Comparing this reduction with the case of increasing the number of RIS elements in the Fig. \ref{fig:N}, we observe that having a larger RIS size has the same effect as increasing BS total transmit power to jointly reduce the used CBLs and improve the network FBL rate. This shows the applicability of our proposed resource allocation scheme in RIS-aided systems over short packet communications to jointly reduce the transmission duration and increase the achievable rate.

    \vspace{-4mm}
    \section{Conclusion}
    \label{conclusion}
    In this paper, we have proposed a multi-objective approach to jointly maximize the total FBL rate among users in a RIS-aided MISO-URLLC system over short packet communications. We identified the total achievable FBL rate in an interference-limited scenario with a penalty term indicating the target error probability impact on the achievable rate. A multi-objective problem has been formulated for maximizing the total FBL rate while minimizing the used CBLs by considering the limited CBL budget to reduce the transmission duration time. The numerical results demonstrated that RIS technology plays a crucial role in URLLC systems in optimizing the total utilized CBLs to realize minimum transmission duration time while maximizing the total rate.


	\bibliographystyle{IEEEtran}
    \vspace{-4mm}
    \bibliography{IEEEabrv,refs}

	
\end{document}